\documentclass[onecolumn,prl,amsmath,amssymb,superscriptaddress,floatfix,preprint]{revtex4}
\usepackage[pdftex]{graphicx}
\usepackage{epstopdf}
\usepackage[pdftex]{graphicx}
\usepackage{epstopdf}
\usepackage{amsbsy} 
\usepackage{xspace}

\usepackage{makeidx}
\usepackage{bbold}
\usepackage{makeidx}
\usepackage{upgreek}
\newcommand{\bstsb}{Bi$_{1.5}$Sb$_{0.5}$Te$_{1.7}$Se$_{1.3}$\xspace}
\makeindex                             
\begin{document}
\title{Conductance spectroscopy of a proximity induced superconducting topological insulator}
\author{M. Snelder}
\affiliation{Faculty of Science and Technology and MESA+ Institute for Nanotechnology, University of Twente, 7500 AE Enschede, The Netherlands}
\author{M. P. Stehno}
\email{m.p.stehno@utwente.nl}
\thanks{Corresponding author}
\affiliation{Faculty of Science and Technology and MESA+ Institute for Nanotechnology, University of Twente, 7500 AE Enschede, The Netherlands}
\author{A. A. Golubov}
\affiliation{Faculty of Science and Technology and MESA+ Institute for Nanotechnology, University of Twente, 7500 AE Enschede, The Netherlands}
\affiliation{Moscow Institute of Physics and Technology, Dolgoprudny, Moscow 141700, Russia}
\author{C. G. Molenaar}
\affiliation{Faculty of Science and Technology and MESA+ Institute for Nanotechnology, University of Twente, 7500 AE Enschede, The Netherlands}
\author{T. Scholten}
\affiliation{Faculty of Science and Technology and MESA+ Institute for Nanotechnology, University of Twente, 7500 AE Enschede, The Netherlands}

\author{D. Wu}
\affiliation{Van der Waals - Zeeman Institute, University of Amsterdam, Sciencepark 904, 1098 XH Amsterdam, the Netherlands}

\author{Y. K. Huang}
\affiliation{Van der Waals - Zeeman Institute, University of Amsterdam, Sciencepark 904, 1098 XH Amsterdam, the Netherlands}
 
\author{W. G. van der Wiel}
\affiliation{NanoElectronics Group and MESA+ Institute for Nanotechnology, University of Twente, 7500 AE Enschede, The Netherlands}
\author{M. S. Golden}
\affiliation{Van der Waals - Zeeman Institute, University of Amsterdam, Sciencepark 904, 1098 XH Amsterdam, the Netherlands}

\author{A. Brinkman}
\affiliation{Faculty of Science and Technology and MESA+ Institute for Nanotechnology, University of Twente, 7500 AE Enschede, The Netherlands}
\date{\today}
\begin{abstract}
We study the proximity effect between the fully-gapped region of a topological insulator in direct contact with an $s$-wave superconducting electrode (STI) and the surrounding topological insulator flake (TI) in Au/\bstsb/Nb devices. The conductance spectra of the devices show the presence of a large induced gap in the STI as well as the induction of superconducting correlations in the normal part of the TI on the order of the Thouless energy. The shape of the conductance modulation around zero-energy varies between devices and can be explained by existing theory of $s$-wave-induced superconductivity in SNN' (S is a superconductor, N a superconducting proximized material and N' is a normal metal) devices. All the conductance spectra show a conductance dip at the induced gap of the STI. 
\end{abstract}
\maketitle
\section{Introduction}
The proximity effect between an $s$-wave superconductor and a three-dimensional topological insulator (3DTI) or nanowire with strong spin-orbit coupling has been predicted to create a Majorana zero-energy mode that may serve as the building block of a topological quantum computer \cite{Kitaev01,Sau2010,Alicea2010,Ivanov01,Sarma2005,Nayak2008,Akhmerov09,qi11,Linder10,Tanaka09,Fu2009,Fu22009,Fu2008}.
The presence of a peak in the conductance at zero-bias has been interpreted as experimental evidence for the existence of Majorana modes in such systems \cite{Deng2012,Mourik2012,Das2012,NadjPerge2014}.
It is also known from theoretical studies that a zero-bias conductance peak is accompanied by conductance dips at the characteristic gap energy which are typical features of Majorana zero-energy modes \cite{Sawa2007,Fu2008,Eschrig,Linder2010,Fu2010,Yamakage2012,Burset2014,Asano2013,Snelder2015}. 

\begin{figure}[t!]
	\centering 
		\includegraphics[width=0.95\textwidth]{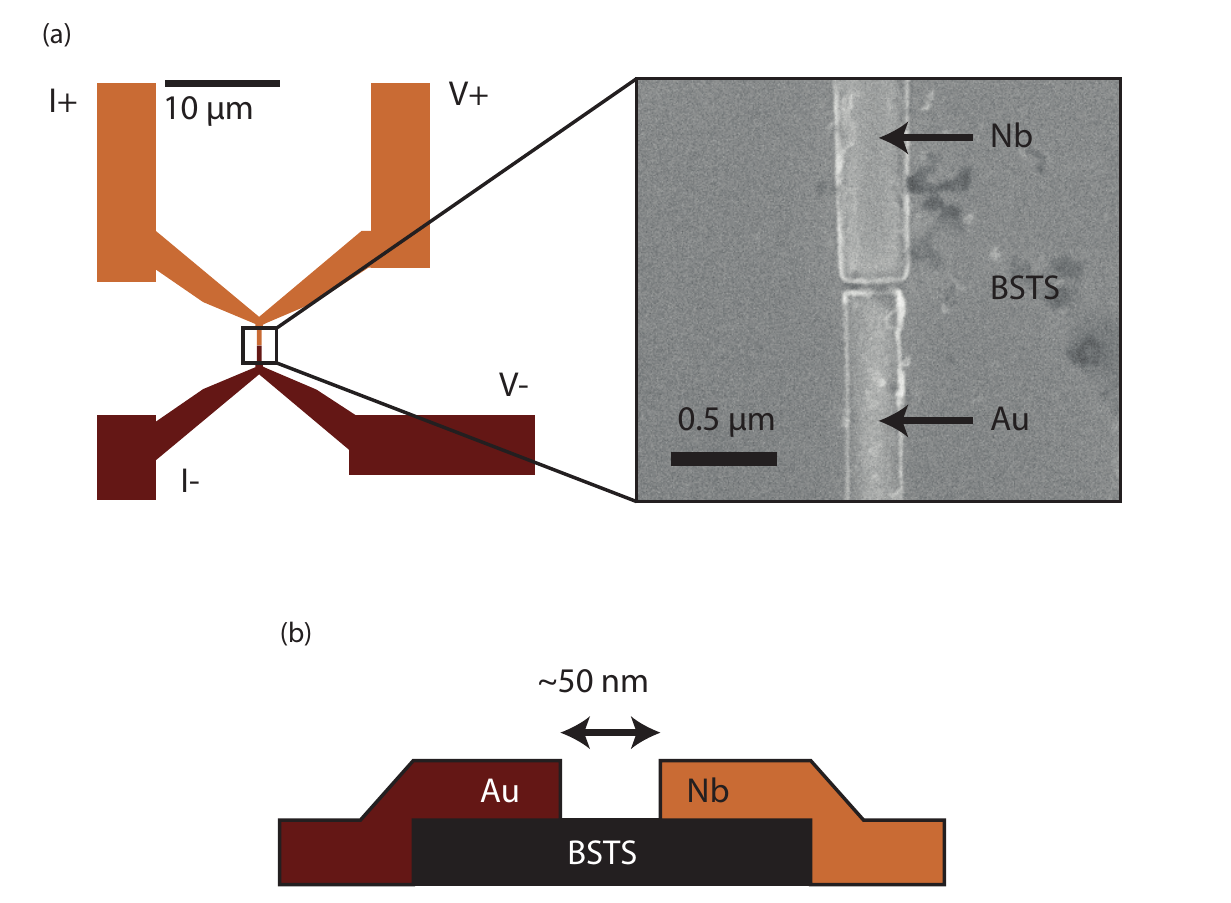}
		\caption{(a) A schematic plan view of the device and SEM image of the small area indicated in the schematic, recorded for a typical Au/\bstsb /Nb trilayer. For clarity, the BSTS flake itself is not shown in the schematic view. Two contacts (left in the sketch) are used for applying a current. The other two contacts (right in the sketch) are used to measure the voltage across the trilayer. The spacing between the Au and Nb leads is 50 nm and the width of the contacts is around 300 nm. (b) Sketch of the side view of the device. The BSTS flakes used had a thickness of about 80 nm.} 
		\label{CH6fig:1}
		\vspace{-15pt}
\end{figure}

Here, we study the proximity effect between an $s$-wave superconductor~({Nb}) and a 3DTI with dominant surface transport~(\bstsb, BSTS) in more detail. Although no Majorana zero-energy modes are expected in these devices, it has been predicted that unconventional superconducting correlations should already be present \cite{Tkachov2013,Tkachov2013b,Snelder2015} at the interface between a topological insulator and an $s$-wave superconductor.
We performed differential conductance measurements on Nb/BSTS/Au devices (see Fig. \ref{CH6fig:1}).
Here, the normal metal reservoir~({Au}) was placed close to the superconductor-3DTI interface with a lateral spacing comparable to the mean free path in BSTS, so as to obtain information about the induced pairing in the 3DTI material.
  
All devices were prepared by exfoliation from a single crystal of \bstsb, which was found to exhibit vanishing bulk conductance in earlier experiments \cite{Snelder2014,Pan2015}. 

We show that the conductance spectra of the devices show the presence of a large induced gap in the STI as well as the induction of superconducting correlations in the normal part of the TI on the order of the Thouless energy.

\section{Conductance measurements}
\begin{figure}[t!]
	\centering 
		\includegraphics[width=1\textwidth]{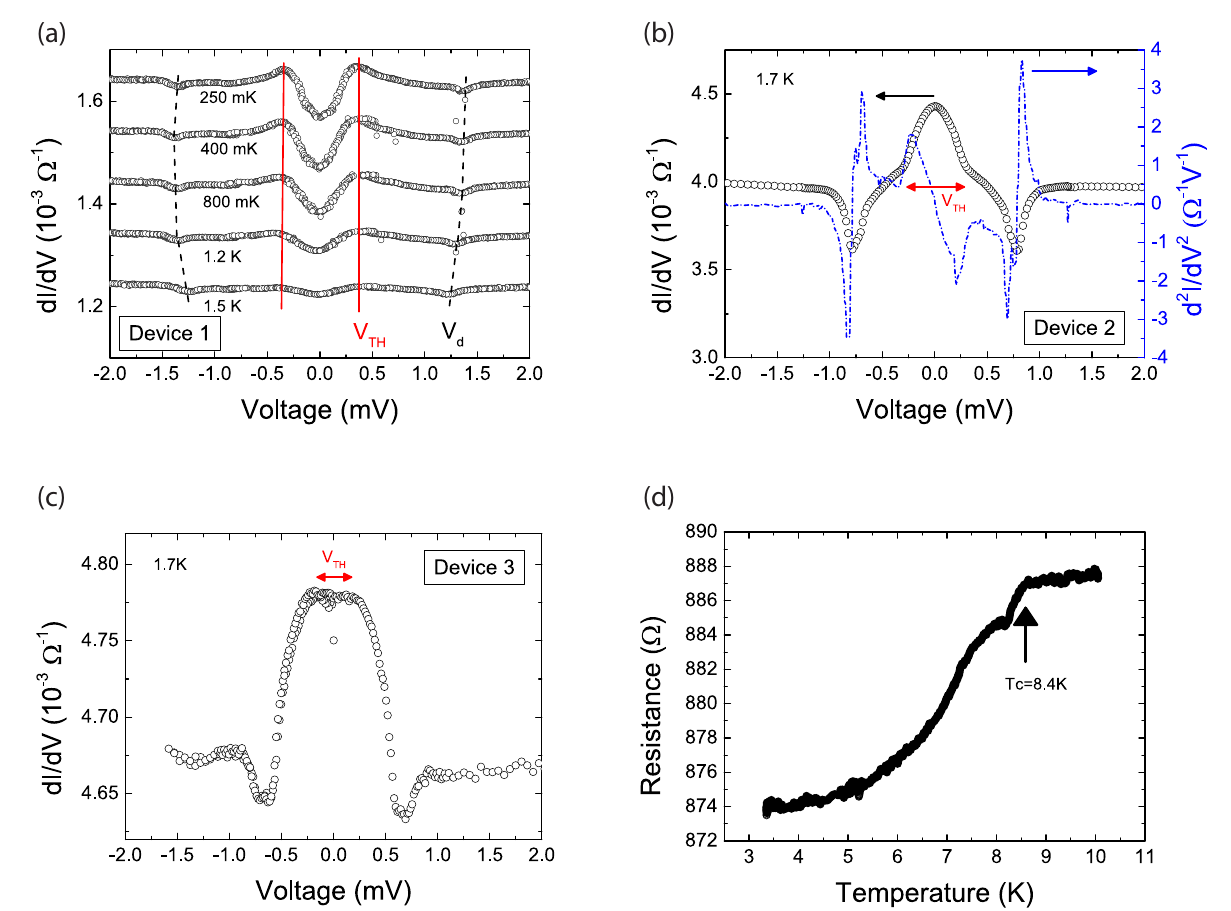}
		\caption{Measurements of Au/\bstsb /Nb devices. (a) Measurements of the conductance of device 1 at different temperatures (for T $<$ 1.5 K, offset in steps of 0.1$\cdot 10^{-3}\Omega$ for clarity). A clear dip is seen in the conductance at energies corresponding to the induced superconducting gap, labelled V$_{d}$. At higher temperatures, $V_{d}$ decreases slightly, as indicated by the black dashed lines.  
The feature at zero energy becomes less pronounced at higher temperature, consistent with the behaviour of the proximity effect for normal $s$-wave superconductivity, and is related to a characteristic Thouless energy scale, marked as V$_{TH}$, which is plotted in Fig. \ref{CH6fig:3}(b). The conductance dips occurring below the gap value of the Nb are reproduced in all three devices. (b) The black symbols show the conductance and blue dot-dashed line shows the derivative of the conductance as function of voltage for device 2. These data show a pronounced dip at a little above $|V|$ values of 0.75 and a smaller conductance dip present at $\pm$1.3 meV. Around zero voltage for device 2 a conductance peak is observed, again related to V$_{TH}$. (c) In device 3, a conductance peak around zero voltage is observed, with a small conductance dip modulation on top. (d) An $R$ vs. $T$ measurement of device 1, showing the determination of the T$_c$ of 8.4 K.} 
		\label{CH6fig:2}
		\vspace{-15pt}
\end{figure}

Two sets of differential conductance measurements were performed with a lock-in amplifier using an AC excitation at 17.7 Hz of 12 nA (device 1) and 50 nA (device 2 and 3). Device 2 and 3 were fabricated simultaneously on a single substrate with flakes that originated from the same cleavage. Device 1 was prepared separately with a flake from a different cleavage cycle. Metal electrodes were patterned using standard electron-beam lithography techniques and sputter deposition. Conductance spectra of device~1 (depicted in the inset of Fig.~\ref{CH6fig:1}(a)) were recorded at several temperatures in the range between 250~mK and 1.5~K  (see Fig.~\ref{CH6fig:2}(a)).
Devices~2 and~3  were measured at a fixed temperature of 1.7~K (Figs.~\ref{CH6fig:2}(b) and~(c)).
Although the line shapes of individual devices look distinctly different, several common features are readily identified: all traces display a dip in the conductance at voltages comparable to the gap energy in the superconductor.
This is clearly seen in device~1 at the lowest measurement temperature where the recorded dip voltage $V_d = 1.3$~mV is close to the expected gap of the Nb electrode with critical temperature $T_c = 8.4$~K \footnote{T$_{c}$ is measured as an abrupt drop in the device resistance
in an R vs. T measurement, as shown in Fig.\ref{CH6fig:2}(d).}  
As the measurement temperature is increased, the position of the conductance dip moves toward lower voltage.
The temperature dependence of the dip position $V_d$ of device 1 is shown in Fig.~\ref{CH6fig:3}(b). Also the dip positions of devices 2 and 3 are plotted at 1.7 K. In devices~2 and~3 the dip features appear to have moved significantly below the bulk gap of niobium.
 However, in device~2 a small depression remains at 1.3~mV which is seen more clearly in the derivative of the conductance spectrum (dashed trace in Fig.~\ref{CH6fig:2}(b)).
 
The presence of a second, lower characteristic energy scale is evident from the modulation of the conductance spectra at smaller voltages.
Devices~1 and~3 show a dip, whereas device~2 exhibits a zero-bias conductance peak.
The widths of the features range between 400 and 800~$\upmu V$ and are indicated by $V_{TH}$ in Fig. \ref{CH6fig:2}(a--c).
Unlike the positions of the dips in the conductance spectra of device~1, the voltages at which the peaks appear do not show a strong temperature dependence, as can be seen from the dark blue symbols in Fig.~\ref{CH6fig:3}(c). 

\section{Discussion} 
\subsection*{Superconducting proximity effect}
Below, we will argue why the measurement data can be interpreted as a strong proximity effect between the Nb superconducting electrode and the underlying TI, which transforms the TI into a superconducting topological insulator (STI) with a sizeable induced gap. 
The STI, then, has a lateral proximity effect with the portion of the TI flake that is not covered by the Nb electrode
(see Figs. 1b and 3a), and this gives rise to peaks at the Thouless energy for a diffusive disordered conductor.

\begin{figure}[t!]
	\centering 
		\includegraphics[width=1\textwidth]{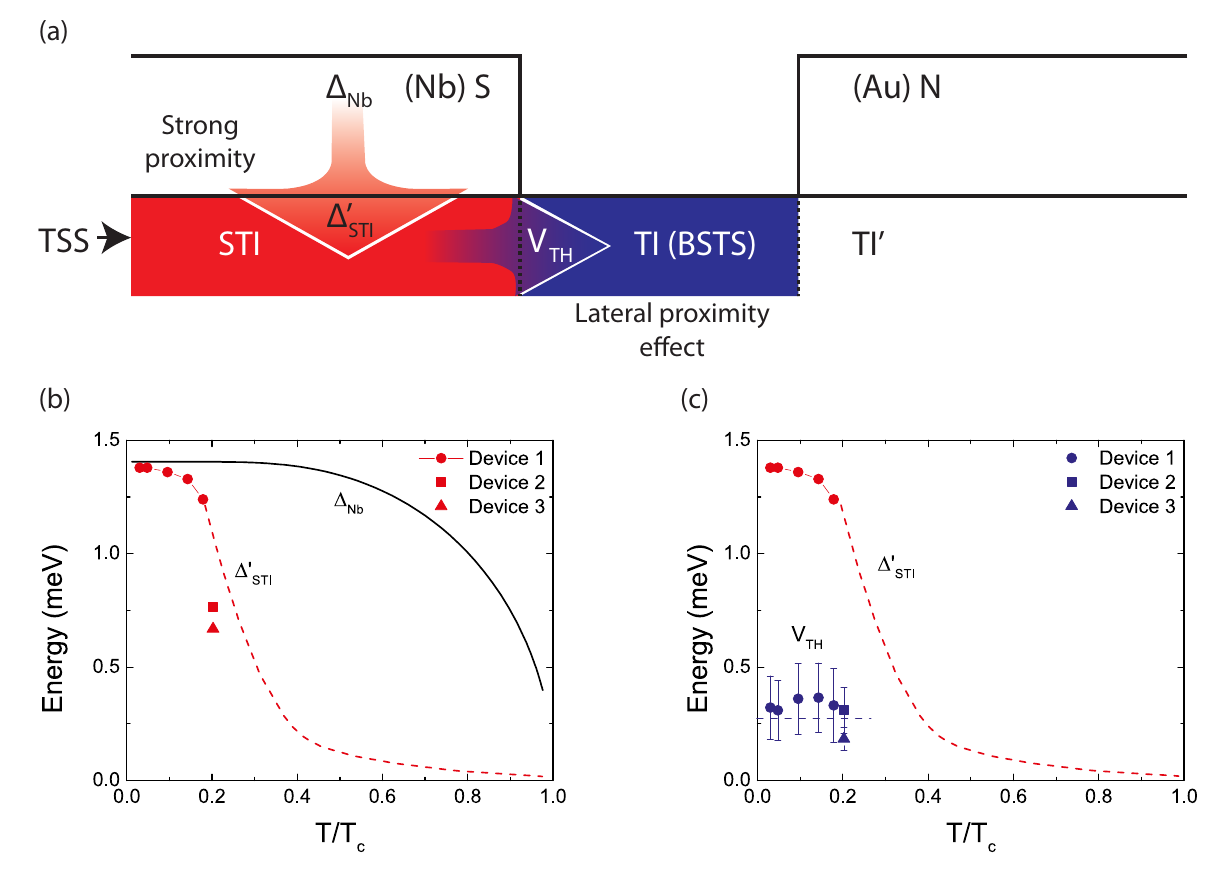}
		\caption{(a) Schematic of the two proximity-based phenomena in Au/\bstsb/Nb devices. The Nb induces a superconducting gap into the topological surface states (TSS) of the topological insulator directly underneath via a strong proximity effect. This creates a superconducting TI (STI, indicated in red), which in turn induces superconducting correlations laterally into the uncovered part of the TI flake (indicated in blue). (b) The Nb $s$-wave BCS superconducting gap (black, solid line) and the induced gap $\Delta'_{STI}$ in the TSS as a function of temperature. Data for all three devices reported are shown, and the red dashed line serves as a guide to eye. (c) The temperature dependence of the induced gap $\Delta'_{STI}$ is contrasted to the relative insensitivity to temperature of the energy scale of the laterally-induced Thouless proximity feature in the TI. The dashed-blue line serves as a guide to the eye.} 
		\label{CH6fig:3}
		\vspace{-15pt}
\end{figure}

Theoretical studies on the influence of the superconducting proximity effect by means of Green's function techniques make use of two important parameters: $\gamma$ and $\gamma_{B}$, defined as
\begin{eqnarray}
\gamma &=&\dfrac{\rho_{S}\xi_{S}}{\rho_{N}\xi_{N}}, \hspace{0.5cm} \gamma_{B}= \dfrac{R_{B}}{\rho_{N}\xi_{N}}.
\end{eqnarray}
The former is a measure of the strength of the proximity effect and the latter is related to the transparency of the barrier.
Here, $R_{B}$ is the interface resistance of the superconductor/TI boundary, $\rho_{S,N}$ and $\xi_{S,N}$ are the normal resistivities and coherence length of the superconductor (S) and topological insulator (N) material, respectively. The induced gap $\Delta'_{STI}$ due to the proximity of Nb to the TI shown in Fig. \ref{CH6fig:3}(a) and (b) is determined by the values of $\gamma$ and $\gamma_{B}$ at the Nb/BSTS interface. For a two-dimensional electronic system in contact with a superconductor, $\gamma$ is small because of the small ratio of the resistivities (as the topological surface states are 2D and the Nb superconductor 3D), and an upper limit of 0.01 can be used \cite{Neurohr}. Due to this small value for $\gamma$ the inverse proximity effect can be neglected and, as a result, the gap size of Nb is unaffected and decoupled from the STI below \cite{Volkov1994,Volkov1994b,Golubov1989,Golubov1995,Aarts1997}.
The temperature dependence of the Nb gap expected from BCS theory is shown as a black, solid line in Fig. \ref{CH6fig:3}(b).
However, the observed dip feature in device 1 (red circles) decreases faster as function of temperature.
We attribute this feature to the induced gap $\Delta'_{STI}$.
In Ref. \cite{Golubov1995}, for example, it is shown that the induced gap (dashed curve in Fig. \ref{CH6fig:3}(b)) decreases faster as function of temperature compared to BCS theory (solid curve in Fig. \ref{CH6fig:3}(b)) although it has the same critical. At low temperatures (device 1), the dip feature reaches 1.3 mV which is close to the Nb gap. 
This indicates that the interface between the Nb and BSTS has high transparency.
The energy corresponding to the dip features in devices 2 and 3 are also plotted in Fig. \ref{CH6fig:3}(b), and these can be seen to be lower in energy than the dip feature of device 1, which is most probably due to a less transparent interface for devices 2 and 3.

As mentioned above, besides this perpendicular proximity effect, there also exists a lateral proximity effect between the STI and the uncovered portion of the topological insulator (shown in blue and labelled TI (BSTS) in Fig. \ref{CH6fig:3}(a)) \cite{Volkov1994,Volkov1994b,Golubov1989,Golubov1995,Aarts1997}. In both the ballistic and the diffusive limit the characteristic energy scale associated with the induced superconductivity is related to the dwell time in the junction and is known as the Thouless energy (data-points shown in Fig. \ref{CH6fig:3}(c)).
For the ballistic limit it is given by $E=\hbar v_{f}/L$ where $v_{f}$ is the Fermi velocity in the normal metal and $L$ the system size. In the diffusive limit it is defined as $\hbar D/L^{2}$ where $D$ is the diffusion constant \cite{Volkov1993,Belzig1996,Melsen1996}. The expression for the ballistic limit gives a value of 2.2 meV whereas we obtain for the diffusive limit an energy value between 0.15 and 0.6 meV. We used a Fermi velocity of 4.5$\cdot$10$^{5}$ m/s \cite{Taskin}, which is in keeping with ARPES data recorded from the same batch of BSTS single crystals as used to make the flakes used in this study \cite{Frantzeskakis2015}, a mean free path between 10 and 40 nm \cite{Snelder2014} and a junction dimension of 100 nm.
The value for the ballistic limit exceeds the energy associated with the gap of Nb so that no features due to ballistic transport are expected below the energy gap.
However, the diffusive Thouless energy corresponds very well to the energy scale $V_{TH}$.
This suggests that mainly diffusive transport contributes to the measured conductance spectra and that the observed features can be interpreted as signatures of laterally induced superconductivity in BSTS.
To conclude, we see that $\Delta'_{STI}$ and the induced superconducting correlations in the normal TI part of the BSTS flake show similar behaviour in all three devices. 

\subsection*{The shape of the conductance spectra} 
We interpreted the modulation of the conductance near zero energy as a signature of induced superconductivity in the normal TI part of the BSTS flake.
The common energy scale associated with the features is at the order of the Thouless energy.
Despite the similar energy scale, comparison of Figs. 2a, 2b and 2c shows the shape of the d$I$/d$V$ curves around zero energy look quite different from each other.
In device 1 we see a clear conductance dip, device 2 exhibits a conductance peak and device 3 a broad conductance peak with a small conductance dip modulation on top.
The appearance of conductance peaks and dips at the order of the Thouless energy is extensively studied in Refs. \cite{Volkov1993,Tanaka2003}, and it was shown that for a Thouless energy smaller than the gap value, both conductance peaks and dips near zero-bias can be observed, depending on the ratio of the barrier strengths and the ratio of each barrier to the resistance of the normal part.
We could not determine the contact resistance of the electrodes due to the inhomogeneity of the BSTS (see Supplementary Information).
However, from the relative magnitude of the conductance modulation near zero voltage we extract that the barriers and the resistance of the normal part are of the same order of magnitude \cite{Volkov1993}.
The combination of a resistance ratio of order unity and a small Thouless energy results a high sensitivity to small variations in the resistance ratios \cite{Tanaka2003} in terms of whether a peak or a dip around zero voltage is seen in the conductance. 
Therefore, even when the devices are fabricated using the same procedure, a small variation in the barrier resistance, something that naturally occurs in the growth procedure, can result in the observed differences in the conductance traces around zero voltage \cite{Poirier1997}.
The existence of both zero conductance peaks and dips has also observed experimentally in semiconductor/superconductor \cite{Poirier1997,Yu2014}, graphene/superconductor \cite{Popinciuc2012} and in topological insulator/superconductor devices \cite{Yang2012,Finck2014}.

The conductance dip at $\Delta'_{STI}$ is reproduced in all three devices. There is an additional small conductance dip at the Nb gap (1.3 meV) in device 2 which is more clear in the derivative of the conductance spectrum. We speculate that this feature at the Nb gap energy in device 2 is due to direct proximity between the uncovered TI part and the Nb superconductor. Normally a strong coherent peak is expected at the (induced) gap energy resulting in a conductance peak instead of a dip.
We will discuss the origin of this conductance dip in the remaining part of this section.

A scenario which results in a conductance dip near the gap energy is the presence of charge imbalance but we rule that out due to the absence of a resistance peak at $T_{c}$ in the $RT$ measurement \cite{CaddenZimansky2007,Clarke1972,Tinkham1972,Schmid1975}.

It is known that the proximity effect can change the local density of states (LDOS) in the normal layer drastically.
In Refs. \cite{Reutliner2014,Belzig1996,Hammer2007,Heikkila,Wilhelm2000,Golubov,Bezuglyi2005} it is shown theoretically that a suppression of the LDOS is possible.
In Ref. \cite{Reutliner2014} this phenomenon is explained by the properties of the contacts and the existence of Andreev bound states at energies just below the gap energy.
These Andreev bound states are related to the tunneling character of the contacts, i.e. contacts with a transmission close to zero.
If the contacts have a significantly higher transparency, the contribution of these bound states to the total measured conductance decreases, resulting in a suppression of the LDOS. A small $\gamma$ can enhance this feature as shown in Ref. \cite{Golubov} and in the Supplementary Information. 
\\
\\
It is an open question whether the dips can be explained by $s$-wave theory or whether a {\it combination} of $s$ and $p$-wave correlations is needed. From previous research it is known that in the time-reversal symmetric case, an equal admixture of $s$ and $p$-wave correlations exist at the surface \cite{Tkachov2013,Tkachov2013b,Snelder2015}.
It has also been shown by Tkachov and Hankiewicz \cite{Tkachov2013,Tkachov2013b} that the $p$-wave correlations decay faster as function of distance than the $s$-wave correlations.
Because the mean free path in our samples is of the same order as the electrode spacing, we believe we are in the right regime to observe $p$-wave correlations if they exist in our devices.
Existing $s$-wave theory on diffusive SNN' devices can be very well applied to the modulation of the conductance in our devices around zero-bias.
The presence of a conductance dip at the induced gap, is - however - less trivial to explain.
We excluded charge imbalance as a cause of the conductance dips.
We know that conductance dips at the induced gap is one of the signatures of $p$-wave correlations. The discussed $s$-wave models in the Supplementary and in Refs. \cite{Reutliner2014,Belzig1996,Hammer2007,Heikkila,Wilhelm2000,Golubov,Bezuglyi2005} make use of Kupriyanov-Lukichev boundary conditions.
A small total resistance implies automatically,  therefore, that the interface is highly transparent.
In the case of a transparent interface, the suppression of the LDOS near the NN' interface is not observable in the conductance as the contribution of this interface to the total conductance can be neglected (see Eqs. (15) and (14) in Refs. \cite{Tanaka2003,Volkov1993} respectively).
Moreover, in the simulations in Ref. \cite{Reutliner2014} a hard wall condition is assumed at the NN' interface (which in this case is the TI/Au interface), which is not applicable to our devices due to the large contact area involved.
In the context of the data presented here, a new model is needed for an SNN' device, that enables the elucidation of the regime in which the NN' interface can have intermediate transparency values and $p$-wave correlations can exist in the N-layer.
As a starting point, the model in Ref. \cite{Tanaka2003} could be extended to use circuit theory boundary conditions at the NN' interface.
From this model we can extract under which conditions the conductance dips still exist when we move slowly away from hard wall conditions and hence investigate if such a feature can still be present when the barrier resistance is small but with finite transmission or that it is present when a specific percentage of $p$-wave correlations is taken into account in the N-layer.      \\
\\ 
 
\section{Methods}
All devices were based on material from a single crystal of BSTS. BSTS crystals were grown using the Bridgman method as described briefly in Refs. \cite{Snelder2014,Pan2015}. Flakes were transferred from a BSTS crystal to a Si/SiO$_{2}$ substrate by means of mechanical exfoliation. Device~2 and~3 were fabricated simultaneously on a single substrate with flakes that originated from the same cleavage. Device~1 was prepared separately with a flake from a different exfoliation cycle. By means of e-beam lithography (EBL), the Au electrode geometry is defined on the flake. A low voltage 15 s etching step is carried out to avoid significant damage of the surface, followed by sputtering deposition in-situ of a 3 nm Ti adhesion layer and a 60 nm Au contact layer. Next, a second EBL step is performed to define the Nb part of the device. After the EBL step, the structure is etched for 15 s at low voltage, followed by in-situ sputtering of 80 nm Nb. BSTS has a small mean free path of 10--40 nm, so the EBL step was optimised to allow inter-electrode spacings of only 50 nm, so as to ensure full proximisation of the BSTS by the superconductor \cite{Snelder2014}. The width of the electrodes is about 300 nm (see Fig. 1a). To avoid damage to the material by physical or chemical etching, we omitted shaping the 3D-TI flakes. Therefore, a small contribution to the device conductances results from current spreading laterally on the surface of the flake. Based on the device dimensions, we estimate this contribution to be $\sim 10\%$ for devices~2 and 3 and $\sim 20\%$ for device~1. We estimated the total resistance of the surrounding BSTS and the resistance of the BSTS between the electrodes by determining the width and length ratio, and applying a parallel resistor model assuming a constant resistivity.

\section{Acknowledgements}
We would like to thank Dick Veldhuis and Frank Roesthuis for technical support during measurements and fabrication. This work is supported by the Netherlands Organization for Scientific Research (NWO), by the Dutch Foundation for Fundamental Research on Matter (FOM), the European Research Council (ERC) and supported in part by Ministry of Education and Science of the Russian Federation, grant no. 14Y26.31.0007. 
\section{Author contributions statement}
AB and MSG supervised the project. MS fabricated, measured and analyzed the devices. MPS, CGM and TS measured and analyzed the devices. AB, MS, MPS wrote the manuscript text incorporating feedback from MSG. AAG developed the code of the simulations presented in the supplementary. DW and YKH grew the BSTS crystals. WGW supported the experiments performed in the He-3 refrigerator.  
\section{Additional information}
The authors declare no competing financial interests. 
\end{document}


\title{Supplementary: Conductance spectroscopy of a proximity induced superconducting topological insulator}
\author{M. Snelder}
\affiliation{Faculty of Science and Technology and MESA+ Institute for Nanotechnology, University of Twente, 7500 AE Enschede, The Netherlands}
\author{M. P. Stehno}
\email{m.p.stehno@utwente.nl}
\thanks{Corresponding author}
\affiliation{Faculty of Science and Technology and MESA+ Institute for Nanotechnology, University of Twente, 7500 AE Enschede, The Netherlands}
\author{A. A. Golubov}
\affiliation{Faculty of Science and Technology and MESA+ Institute for Nanotechnology, University of Twente, 7500 AE Enschede, The Netherlands}
\affiliation{Moscow Institute of Physics and Technology, Dolgoprudny, Moscow 141700, Russia}
\author{C. G. Molenaar}
\affiliation{Faculty of Science and Technology and MESA+ Institute for Nanotechnology, University of Twente, 7500 AE Enschede, The Netherlands}
\author{T. Scholten}
\affiliation{Faculty of Science and Technology and MESA+ Institute for Nanotechnology, University of Twente, 7500 AE Enschede, The Netherlands}
\author{D. Wu }
\affiliation{Van der Waals - Zeeman Institute, University of Amsterdam, Sciencepark 904, 1098 XH Amsterdam, the Netherlands}
\author{Y. K. Huang}
\affiliation{Van der Waals - Zeeman Institute, University of Amsterdam, Sciencepark 904, 1098 XH Amsterdam, the Netherlands}
\author{W. G. van der Wiel}
\affiliation{Faculty of Science and Technology and MESA+ Institute for Nanotechnology, University of Twente, 7500 AE Enschede, The Netherlands}
\author{M. S. Golden}
\affiliation{Van der Waals - Zeeman Institute, University of Amsterdam, Sciencepark 904, 1098 XH Amsterdam, the Netherlands}
\author{A. Brinkman}
\affiliation{Faculty of Science and Technology and MESA+ Institute for Nanotechnology, University of Twente, 7500 AE Enschede, The Netherlands}
\date{\today}
\maketitle

\section*{Contact resistance}
To determine the contribution of the contact resistance between BSTS and the electrodes, we prepared Au/BSTS/Au devices using the same growth parameters as described under ``methods" in the manuscript with channel lengths between 2 $\mu$m and 4 $\mu$m and a width of 2 $\mu$m. The channels are prepared on the same flake that is shaped into different channel lengths by means of Ar ion milling. In this way we also avoid ambiguity in the measured voltage drop \cite{Lee2014}. Due to the larger areas of the channels compared to the Au/BSTS/Nb devices used for the data in the main body of the paper, we expect that the additional Ar ion milling step has insignificant influence on the total resistance. The contact area between the electrode and the BSTS is 1 $\mu$m$^{2}$. Thereafter, we performed a two point measurement of the resistance as function of temperature for the various channels in a Physical Property Measurement System (PPMS). 

In Fig. S1(a) a microscope image is shown of such a device to determine the contact resistance between Au and BSTS. Fig. S1(b) shows data obtained from this device. We see that above 200 K the resistance as function of the channel length is linear, as expected. However, for 200 K and lower, the data points start to deviate from the linear behaviour. 

To exclude that a difference in thickness of the different channels is the origin of the resistance behaviour as function of temperature, we measured the thickness of the various parts using an atomic force microscope (AFM). The average thickness was 112 nm with a deviation of only 4 nm, which is too small to explain the large difference in resistance among the channels. 

Instead, the deviation can be understood in terms of the intrinsic inhomogeneity of BSTS due to the high degree of alloying in the system. We fitted the R vs.T curves of the different channels with $\sim e^{-E_{g}/k_{B}T}$ where $E_{g}$ is the distance of the Fermi level to the conduction band. For the channels of 1.5, 2.5 and 3.5 $\mu$m (channel 1, 2 and 3 respectively) we obtained for $E_{g}$ 1.35 meV, 0.85 meV and 1.20 meV, respectively. In a semiconductor picture this means that for lower temperatures more carriers are present per unit area for channel 2, followed by channel 3 and then channel 1.  Considering Fig. S1(b) again, this picture is consistent with the resistance distribution over the three channels. This implies then that the difference in carrier densities in the three channels is due to differences in $E_{g}$, which - although modest - are sufficient to dominate over the dependency on the channel length at low $T$. The modest variation in the position of impurity states able to be thermally excited for transport is probably not unexpected in such a heavily alloyed system as BSTS. 

Although this interpretation is consistent with the data, it does not exclude the possibility that the origin of the deviation of the data from a linear behaviour is due to a difference in contact resistance between the Au electrodes and the BSTS. In that case, the contacts can also behave differently as a function of temperature which will lead to scatter in the data as a function of channel length. 

In both cases, the contribution of the contact resistances to the conductance measurements remains unknown. Therefore, the contact resistance is treated as an additional parameter in the interpretation of the data.

\section*{Suppression of LDOS}
In Fig. S2(a) simulations of the LDOS at the NN' interface of an SNN' device are presented. The corresponding device is depicted in Fig. S2(c). For this purpose, we used the numerical code from Ref. \cite{Brammertz}. We simulated the trilayer model for $\gamma_{1}=0.01$ and different barriers. $\gamma_{1}$ and $\gamma_{2}$ are related to the ratio of the normal state resistivity of the superconductor and the resistivity of the N layer and to the ratio of the resistivity of the N and N' layer, respectively (see Fig. 3a of the main text). $\gamma_{B1}$ and $\gamma^{*}_{B2}$ are related to the barrier strength at the SN interface and NN' interface respectively. We see that for a lower value of $\gamma_{B1}$, i.e. higher transparency of the Nb/BSTS interface, the suppression of the LDOS increases. The barrier at the Au/BSTS interface or $\gamma^{*}_{B2}$ is less relevant. A decrease in $\gamma_{2}$ means a smaller resistivity of the N layer compared to the N' layer, resulting in an increased proximity effect and therefore a larger induced gap. This enhancement implies that there is a stronger coupling to the superconductor and therefore a decrease in tunneling character of the Nb/BSTS barrier. Therefore we also see next to the larger induced gap a larger suppression at the gap energy (compare, for example, the solid and the dashed traces in Fig. S2(a)). 

Figure S2(b) shows the spectral transmission coefficient, $T(E)$, of the total device around the gap energy calculated by the same methods as in Ref.\cite{Volkov1993}. Although a clear suppression is visible in the LDOS at the NN' interface, the suppression does not appear in the total spectral transmission of the device. At zero temperature the differential conductance is $\sigma(V)$ is equal to $T(V)$.

\begin{figure}[t!]
	\centering 
		\includegraphics[width=1\textwidth]{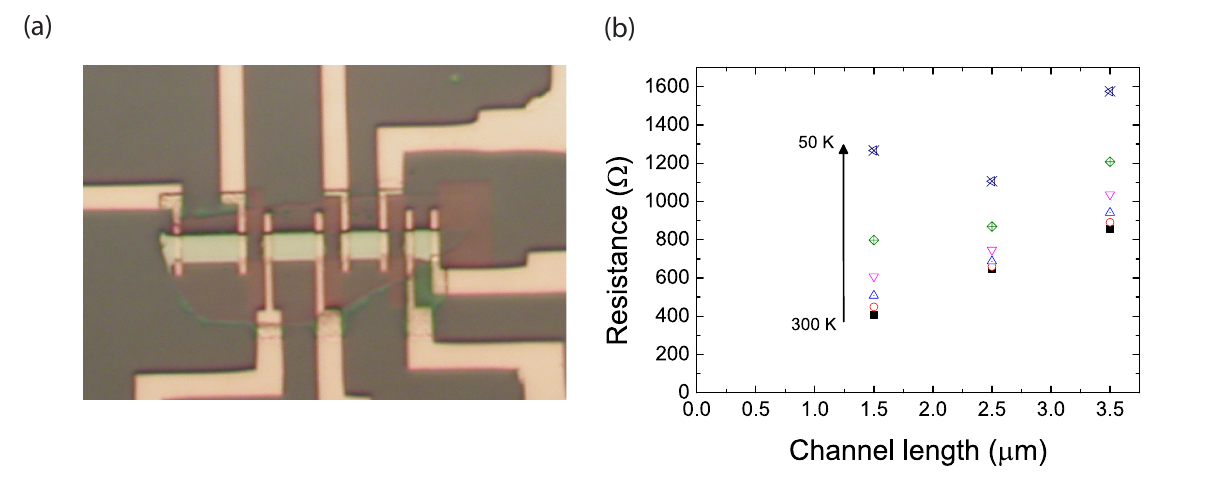}
		{FIG. S1: (a) Micrograph of a device to determine the contact resistance between Au and \bstsb. The contact area is 1 $\mu$m$^{2}$. The smallest channel length is 1.5 $\mu$m. The largest is 4.5 $\mu$m in this picture. Around the leads the original shape of the flake can still be recognized by the colour difference between the middle area of the picture and the surroundings. (b) The resistance of the different channels from 300 to 50 K, measured for T steps of 50 K.} 
		\label{CH6fig:1b}
		\vspace{-15pt}
\end{figure}
\begin{figure}[t!]
	\centering 
		\includegraphics[width=0.95\textwidth]{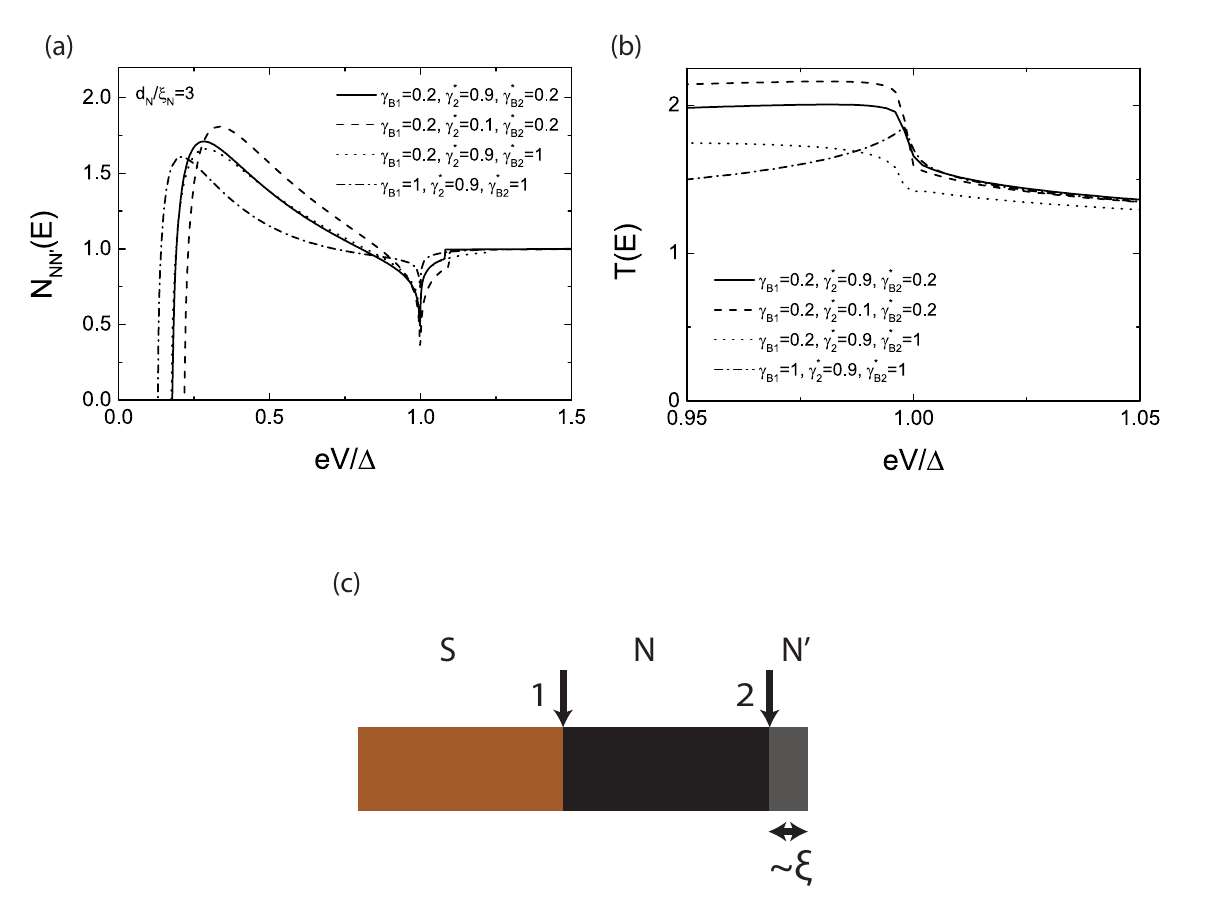}
		{FIG. S2: (a) Simulations of an SNN' device where N has a spatial extent 3 times the superconducting coherence length. For lower barrier height the suppression of the LDOS at the NN' interface become larger. (b) Spectral transmission coefficient of the same schematic device. The suppression of the LDOS at the NN' interface is not visible in the total transmission coefficient. (c) Schematic of the simulated device. The numbers 1 and 2 indicate the interfaces. The N' layer in the model has a spatial extent of order of the coherence length.} 
		\label{CH6fig:4}
		\vspace{-15pt}
\end{figure}